\documentclass[aps,singlecolumn,showpacs,superscriptaddress]{revtex4}
\usepackage{amsmath,amssymb}
\usepackage{graphics,graphicx}
\usepackage{dcolumn,bm}
\usepackage{psfrag}
\begin{document}

\title
{Annual Journal citation indices: a comparative study}

\author{Abdul Khaleque}
\email[Email: ]{aktphys@gmail.com}
\affiliation{Department of Physics, University of Calcutta,
92 Acharya Prafulla Chandra Road, Kolkata 700 009, India}

\author{Arnab Chatterjee}
\email[Email: ]{arnabchat@gmail.com}
\affiliation{Condensed Matter Physics Division, Saha Institute of Nuclear 
Physics, 1/AF Bidhannagar, Kolkata 700 064, India}

\author{Parongama Sen}
\email[Email: ]{psphy@caluniv.ac.in}
\affiliation{Department of Physics, University of Calcutta,
92 Acharya Prafulla Chandra Road, Kolkata 700 009, India}

\begin{abstract}
We study the statistics of citations made to 
the indexed Science journals 
in the Journal Citation Reports during the period 2004-2013 using 
different measures. We consider different measures which quantify  the impact of the journals.
To our surprise, we find that the apparently uncorrelated measures, even when defined in an arbitrary manner,
show strong correlations. This is checked over all the years considered.  Impact factor being one of these measures, 
the present work raises the question whether it is actually a nearly perfect index as claimed often.
 In addition we study the distributions of the different indices which also behave similarly.

\end{abstract}

\maketitle

\section{Introduction}

Extensive studies have been made to analyze quantitatively the popularity of 
commodities (e.g., books, DVDs), movies, academic publications, webpages etc. over the last two decades or so,
thanks to the availability of such data.  
Remarkably, identical  behavior of the relevant distributions have been observed 
in many cases
suggesting a common dynamical scheme responsible for the universality. To study the popularity of 
a research publication or paper, usually one calculates the citations made to that paper. The citation probability
over time as well as citation  distributions have been studied in great detail in recent years ~\cite{Sen:2013,Shockley:1957,
Laherrere:1998,Redner:1998,Radicchi:2008,newman2005power,Rousseau:1994,Egghe:2000,
Burrell:2001,Burrell:2002,Petersen:2011, katz2000scale,Eom:2011}.
The popularity of an academic journal may also be similarly quantified using the citation data made to
the papers published therein. 
The total citations received in a year, the  impact factor~\cite{Garfield:1964,Garfield:2006}, 
and the eigenfactor~\cite{Bergstrom:2008} are well-known popular measures.
The impact factor (IF)~\cite{Garfield:1964,Garfield:2006} of  
an academic journal is a measure which reflects 
the average number of citations to recent articles published in the same journal. 
It is frequently used as a proxy for the relative importance of a journal within its field, 
with journals with higher IFs deemed to be more important compared to those with lower ones. 
However, according to~\cite{leydesdorff2011integrated}, IF may not be the perfect measure to
compare the quality of two journals.
The \textit{eigenfactor} measure in addition takes into account the quality of the 
journals in which the citing articles appear, arguing that a journal is considered 
to be more influential if it is cited often by other influential journals. 
It was shown~\cite{Fersht:2009} however that the eigenfactor measurement 
is more or less correlated with the annual citation measure.

Apart from studying the properties/statistics of the standard measures of annual citation
and impact factor, we also introduce and analyze a new measure
called the citation rate, defined in the next section.

In the present paper, we analyze the inter-dependence of the three indices, 
correlations of the same measures over time (auto-correlations), 
as well as their distributions. In section 2, we define the quantities considered: the details of the data and results are presented
in section 3 and in the last section summary  and discussions are made.

\section{Definition of citation indices}
Impact factors are calculated yearly for journals that are indexed in the Journal Citation Reports~\cite{JCR}.
The precise definition of IF is the following: if papers published in a journal
in years $T-2$ and $T-1$ are cited ${\mathcal {N}}(T-2) + {\mathcal {N}}(T-1)$ times by indexed journals in the year $T$, 
and $N(T-2) + N(T-1)$ be the number of citable articles published in those years, then 
the impact factor in year $T$ is given by
\begin{equation}
\label{eq:if}
I(T) = \frac{{\mathcal {N}}(T-2) + {\mathcal {N}}(T-1)}{N(T-2) + N(T-1)}.
\end{equation}
One can also measure $n(T)$, the  number of annual citations (AC)  to a journal in a given year.
This is given by 
\begin{equation}
\label{eq:cita}
n(T) = \sum_{t \leq T}{\sum_i {\mathcal{A}}_i (t,T)},
\end{equation}
where ${\mathcal {A}}_i(t,T)$ is the citations received in the year $T$ 
by the $i$ th paper published 
in the year $t \leq T$.

We calculate another index, $r(T)$, the \textit{annual citation rate} (CR)
  at a particular year $T$ 
that is   defined as the number of citations received in a year (annual citations) 
divided by the number of articles published in the \textit{same} year. Formally, 
\begin{equation}
\label{eq:cita-rate}
r(T) = n(T)/N(T).
\end{equation}
Note that this is clearly different from the average citation rate defined in~\cite{ESI}
which denotes average number of citations received in a particular time interval
by all previously published papers. However it is rather arbitrary as the numerator and the denominator
are uncorrelated. We introduce this measure with the purpose to see how important is this arbitrariness.

These three measures are available from a single year's report citation data. Combining data of different years, 
we consider another index $r\prime$ which may be less arbitrary than $r$ but still quite different from $I$. We define $r\prime$ as
$r^{\prime}(T) = n(T)/\langle 
N \rangle$, where $\langle N \rangle$ is the average of $N(T)$ over a extended time interval (10 years in our case).

The number of annual citations $n(T)$ might depend on the age of the journal as 
well as on the number of papers published in it. 
Detailed studies on citation data have shown that a paper's citation 
probability 
decays with time as a power law roughly up to 20 
years after its publication after which it falls 
drastically~\cite{Redner:2005}. So, one can assume that the total citation 
$n(T)$  consists of citation to papers not more than $\sim 20$ years old 
practically. Hence, if the ages of the journals considered are greater than 
$\sim 
20$ mostly, $n(T)$ approximately covers the same time period for all journals 
and age of a journal will not be an important factor. However, $n(T)$ may be 
biased by the number of publications and thus it is meaningful to scale it by a 
typical number of publications (as done for $r$ and $r\prime$). 

\section{Data and Results}
We collected data for all Science journals indexed in ISI Web of Knowledge
for the Science database, containing the following information:
(i) the number of citations $n(T)$ received by the journal in a year $T$
(ii) IF $I(T)$ in that year $T$,
(iii) number of papers $N(T)$ published in that particular year $T$.
The data is for  $10$ years ($2004-2013$) taken from Journal Citation Reports 
(JCR)~\cite{JCR}.

\subsection{Correlations}
\label{sec:correlation}

\begin{figure}[h]
  \begin{center}
  \includegraphics[width=11.0cm]{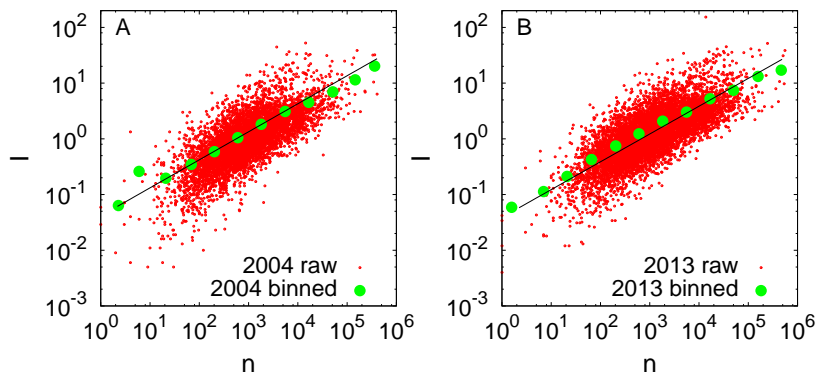}
    \end{center}
      \caption{
    Scatter plot of impact factor (I) vs. citation ($n$). 
  The binned data is also shown, which seems to fit reasonably 
  to $I \propto n^{\xi_n}$. (A) The exponent $\xi_n \approx 0.50$ for 2004
  and (B) $\xi_n \approx 0.49$ for 2013.
  }
\label{fig:nvsi}
 \end{figure}
\begin{figure}[t]
  \begin{center}
  \includegraphics[width=11.0cm]{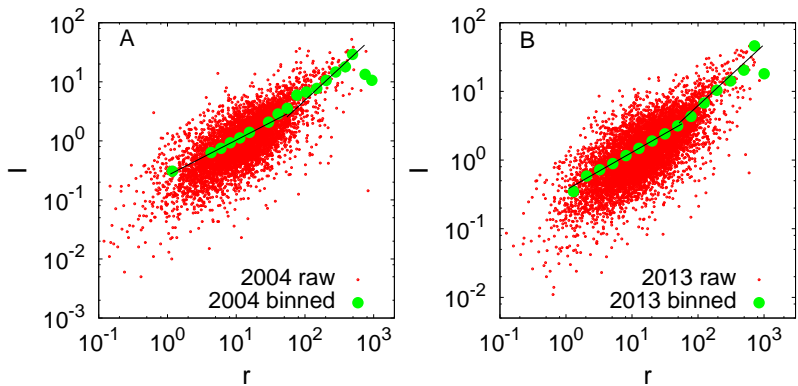}
    \end{center}
      \caption{
    Scatter plot of impact factor (I) vs. citation rate ($r$). 
  The binned data is also shown, which seems to fit 
  to $I \propto r^\xi_r$, with two different exponents. 
  (A) For 2004, $\xi_{r1} \approx 0.60$ for lower $r$
  value and $\xi_{r2} \approx 1.08$ for larger $r$ value.
  (B) For 2013, $\xi_{r1} \approx 0.55$ for lower $r$
  value and $\xi_{r2} \approx 0.89$ for larger $r$ value.
  }
\label{fig:rvsi}
 \end{figure}
\begin{figure}[t]
   \begin{center}
  \includegraphics[width=11.0cm]{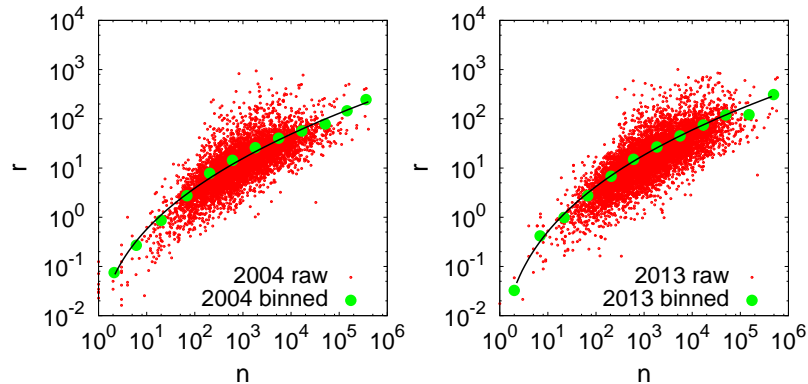}
     \end{center}
      \caption{
    Scatter plot of impact factor ($r$) vs citation ($n$). 
  The binned data is also shown, which fits  well 
  to $r = \exp(c_n+a_n(\log{n})^{b_n})$.  (A) For 2004, the exponents are 
$a_n \simeq 4.32 $, $b_n \simeq 0.39 $ and $c_n \simeq -6.59 $.
  (B) For 2013, the exponents are  $a_n \simeq 6.29 $, $b_n \simeq 0.33 $ and 
$c_n \simeq -8.93 $. 
Details of fitting parameters for different years are shown in 
Table.~\ref{tab:exporn}.}
\label{fig:nvsr}
\end{figure}

\begin{table}[h]
\begin{center}
\caption{Table for the value of the Exponents $a$ and $\xi_n$ for different 
years. The fitted form is $I=an^{\xi_n}$.}
\begin{tabular}{| c| c | c|}
\hline
Year    &       $ a $         &        $ \xi_n $     \\
\hline
 $2004$ & $0.04 \pm 0.01$   &  $0.50 \pm 0.02$\\
 $2005$ & $0.04 \pm 0.01$   &  $0.47 \pm 0.01$\\
 $2006$ & $0.05 \pm 0.01$   &  $0.49 \pm 0.01$\\
 $2007$ & $0.05 \pm 0.01$   &  $0.47 \pm 0.01$\\
 $2008$ & $0.06 \pm 0.02$   &  $0.46 \pm 0.02$\\
 $2009$ & $0.06 \pm 0.03$   &  $0.43 \pm 0.02$\\
 $2010$ & $0.06 \pm 0.01$   &  $0.45 \pm 0.02$\\
 $2011$ & $0.07 \pm 0.05$   &  $0.44 \pm 0.03$\\
 $2012$ & $0.05 \pm 0.03$   &  $0.46 \pm 0.02$\\
 $2013$ & $0.04 \pm 0.03$   &  $0.49 \pm 0.02$\\
 \hline
\end{tabular}
\label{tab:expo}
\end{center}
\end{table}
We first report the correlation between the different measures for different years.
Figs.~\ref{fig:nvsi}  and \ref{fig:rvsi} shows the behavior of $I$ versus $n$ 
and $r$
respectively.
The impact factor $I$ shows remarkable correlation with 
the number of citations $n$ for each year.
In fact, the data binned for number of citations 
shows a very good agreement with a power law: 
$I = an^{\xi_n}$, with $\xi_n =0.47 \pm 0.03$ considering all the years.
 The values of the exponents $a$ and $\xi_n$ for different years are given in Table~\ref{tab:expo}.
The binned data in Fig.~\ref{fig:rvsi} indicate that   $I$ and  $r$ are also 
related by a power law but there are apparently two distinct scaling regimes,
roughly below and above $r \approx 50$.
Fitting the data piecewise by power laws, we get 
$I \propto r^{{\xi_r}}$ with 
$\xi_{r1} = 0.60 \pm 0.02$ for $r < 50$ and $\xi_{r2} = 
1.09 \pm 0.03$ otherwise for $2004$
and  $\xi_{r1} = 0.55 \pm 0.02$ for $r < 50$ and $\xi_{r2} = 
0.89 \pm 0.08$ otherwise for $2013$.
The power law exponent for the low $r$ region is less than that in the high $r$ 
-- a trend that is consistent for all years, except that the exponents are 
slightly different  (see Table~\ref{tab:expoir} for all years).
Fig. \ref{fig:nvsr} shows the variation of $r$ with $n$, from where it is quite 
interesting to note  that the 
annual citations   and citation rates have a different functional dependence. 
Here, the variation of $r$ with $n$ fitted well with the form 
$r =\exp \left[ c_n+a_n(\log{n})^{b_n} \right ]$ and the estimated exponents 
are tabulated in Table.~\ref{tab:exporn}. The most relevant exponent $b_n$
has a value roughly around $0.5$ with some variation for different years.
It is interesting to find that annual citation rate $r$ which is an implicit 
function of the annual citations $n$ has a nontrivial dependence.

\begin{table}
\begin{center}
\caption{Table for the value of the Exponents $a$ and $\xi_r$ for different 
years. The fitted form is $I = ar^{\xi_r}$.}
\begin{tabular}{| c| c | c|c|c|}
\hline
Year    &       $ a $  for low $r$       &        $ \xi_{r1} $ for low $r$ & $ 
a $  for high $r$       &        $ \xi_{r2} $  for high $r$   \\
\hline
 $2004$ & $0.26  \pm 0.01$   &  $0.60 \pm 0.02$ & $0.03  \pm 0.02$   &  $1.10 
\pm 0.03$\\
 $2005$ & $0.24 \pm 0.01$   &  $0.66 \pm 0.01$& $0.09  \pm 0.04$   &  $0.93 \pm 
0.09$\\
 $2006$ & $0.28 \pm 0.01$   &  $0.61\pm 0.01$& $0.07  \pm 0.05$   &  $0.95 \pm 
0.13$\\
 $2007$ & $0.28 \pm 0.02$   &  $0.62 \pm 0.01$& $0.06  \pm 0.06$   &  $0.98 \pm 
0.16$\\
 $2008$ & $0.26 \pm 0.03$   &  $0.65 \pm 0.03$& $0.04 \pm 0.02$   &  $1.07\pm 
0.09$\\
 $2009$ & $0.29 \pm 0.02$   &  $0.62\pm 0.02$& $0.11  \pm 0.05$   &  $0.86 \pm 
0.08$\\
 $2010$ & $0.29 \pm 0.02$ & $0.64 \pm 0.03$& $0.06  \pm 0.03$ & $1.00 \pm 
0.09$\\
 $2011$ & $0.31 \pm 0.03$   &  $0.61 \pm 0.02$& $0.11  \pm 0.03$ &  $0.87 \pm 
0.05$\\
 $2012$ & $0.31 \pm 0.02$   &  $0.60 \pm 0.02$& $0.04  \pm 0.01$   &  $1.04 \pm 
0.06$\\
 $2013$ & $0.36 \pm 0.02$   &  $0.55 \pm 0.02$& $0.10  \pm 0.04$   &  $0.89 \pm 
0.08$\\
 \hline
\end{tabular}
\label{tab:expoir}
\end{center}
\end{table}

\begin{table}
\begin{center}
\caption{Table for the value of  $a$, $b$, $c$ for different 
years. The fitted form is $r =\exp \left[ c +a (\log{n})^{b} \right ]$.}
\begin{tabular}{| c| c | c| c|}
\hline
Year    &       $ a $         &       $b$ & $c$    \\
\hline
 $2004$ & $ 4.32 \pm 1.35$ & $0.40 \pm 0.08$ & $-6.59 \pm 1.44$ \\
$2005$ & $ 4.84 \pm 1.66$ & $0.38 \pm 0.08$ & $-7.23 \pm 1.78$ \\
$2006$ & $ 7.26 \pm 1.96$ & $0.29 \pm 0.06$ & $-9.86 \pm 2.01$ \\ 
$2007$ & $ 3.22 \pm 1.14$ & $0.48 \pm 0.10$ & $-5.42 \pm 1.28$ \\
$2008$ & $ 3.27 \pm 0.85$ & $0.48 \pm 0.08$ & $-5.53 \pm 0.91$ \\
$2009$ & $ 2.90 \pm 0.58$ & $0.50 \pm 0.06$ & $-4.87 \pm 0.64$ \\
$2010$ & $ 1.88 \pm 0.48$ & $0.63 \pm 0.08$ & $-3.58 \pm 0.58$ \\
$2011$ & $ 3.67 \pm 0.70$ & $0.45 \pm 0.05$ & $-5.96 \pm 0.77$ \\
$2012$ & $ 1.38 \pm 0.33$ & $0.73 \pm 0.08$ & $-2.88 \pm 0.42$ \\
$2013$ & $ 6.29 \pm 1.19$ & $0.33 \pm 0.04$ & $-8.94 \pm 1.23$ \\
 \hline
\end{tabular}
\label{tab:exporn}
\end{center}
\end{table}

{\it {Auto-correlation}}:
We have also calculated the dynamic correlation of each of the indices ($n$, $I$, $r$)  with itself over consecutive years.
Plotting the values for two different years for the same journal, the auto-correlation is estimated by 
calculating the correlation coefficient.  
The  linear correlation coefficient is a measure of the strength of linear relation 
between two quantitative variables, say $x_i$ and $y_i$.
We use $R$ to denote the sample correlation coefficient:
\begin{equation}
\label{eq:correlation}
R=\frac{\sum_{i=1}^K (x_i-\bar{x})(y_i-\bar{y})}{\sqrt{{\sum_{i=1}^K (x_i-\bar{x})^2}\sum_{i=1}^K(y_i-\bar{y})^2}}
\end{equation}
Where $K$ is the number of individuals in the sample.

\begin{figure}
\begin{center}
\includegraphics[width=11.0cm]{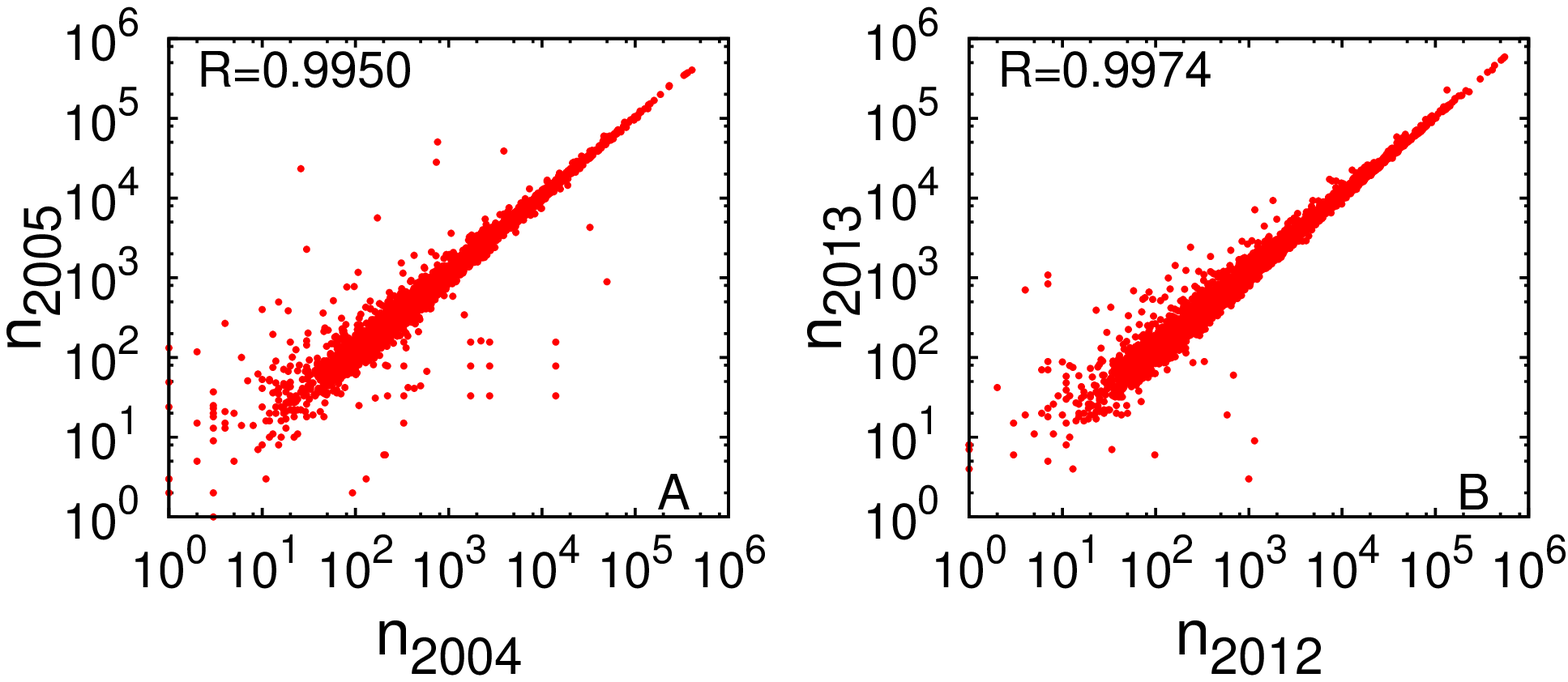}\\
\includegraphics[width=11.5cm]{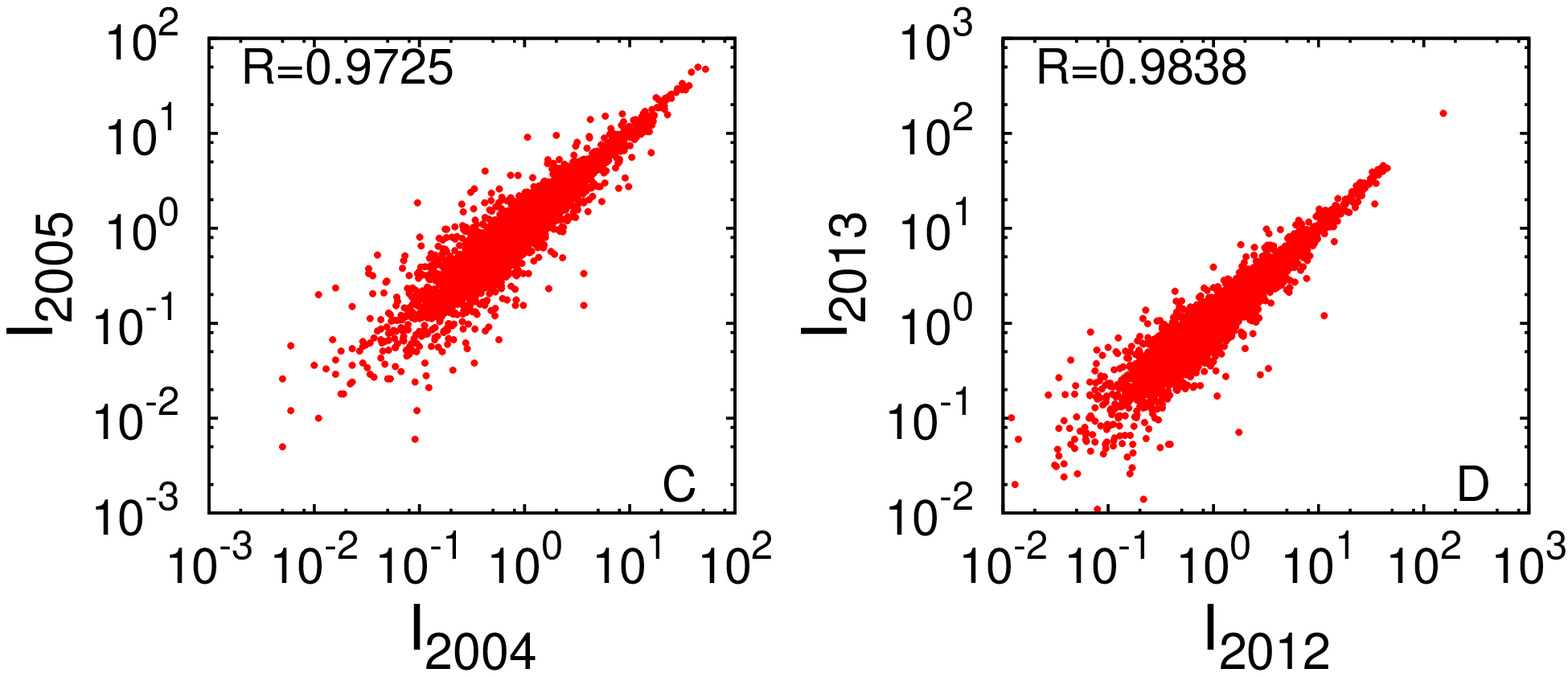}
     \caption{
      Correlation of actual values of annual citations $n$ for two pairs of
     successive years (A) 2004-2005 and (B) 2012-2013. The correlation 
coefficient shows very high values, $0.9950$ and $0.9974$ respectively.
Correlation of actual values of impact factors $I$ for two pairs of
successive years (C) 2004-2005 and (D) 2012-2013. The correlation 
coefficient shows fairly high values, $0.9725$ and $0.9838$ respectively.
}
\label{fig:corractual}
\end{center}
 \end{figure}

In Fig. \ref{fig:corractual}, the correlations for $n$ and $I$ are presented. We choose two consecutive years from the 
extreme ends, i.e., $2004$-$05$ and $2012$-$13$. It is observed that these are highly correlated as $R$ is close to $1$ in  all the cases. 
Such high correlations are apparently not 
present for $r$ for all pairs of consecutive years. In fact, $R$ for $r$ shows considerable fluctuation as shown in Table~\ref{tab:coefficient}.
There may be an upward trend although from the last few data points there seems to be a 
 tendency to stabilize at values which are still not very close to unity 
(compared to the correlation coefficient for $n$ and $I$). 
The lack of strong  correlation in $r$ signifies the fluctuation in the number of publications even for consecutive years.

\begin{figure}
\begin{center}
\includegraphics[width=14.0cm]{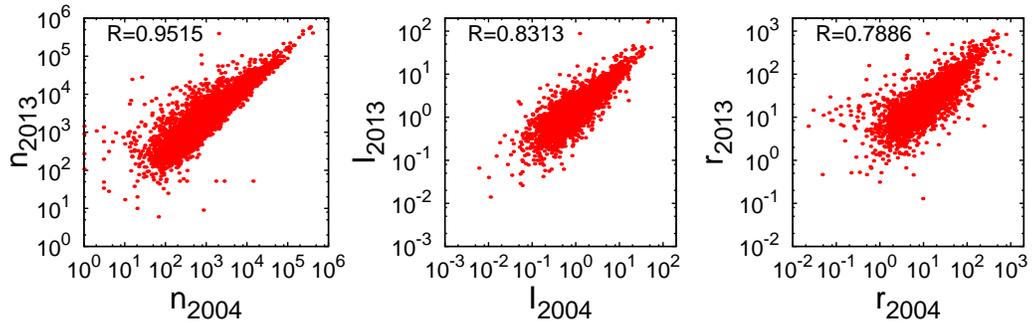}
     \caption{
      Correlation of actual values of (A) annual citations $n$ for the pair of
    years 2004-2013. The correlation 
coefficient is $0.9515$.
(B) Same for impact factors $I$ for the pair of
    years 2004-2013. The correlation 
coefficient is $0.8313$.
(C) Same for citation rates $r$ for the pair of
    years 2004-2013. The correlation 
coefficient is $0.7886$.
}
\label{fig:corractual1}
\end{center}
 \end{figure}

Apart from estimating  the data for consecutive years, we have also calculated the correlation 
for the two extreme years for which data is available.  The correlations for $n$, $I$ and $r$ of two extreme years $2004$ and $2013$ are
shown in Fig.~\ref{fig:corractual1}. The value of $R$ in this case is still close to unity for $n$ while for $I$ and $r$ it is much less.
That $R$ is close to $1$ for $n$ over a comparatively long time interval (9 years) is not surprising as for $n$ citations to all 
previously published papers are counted. On the other hand, for $I$, the correlation drops since citations made for papers published 
two years prior to $2004$ and that to $2013$ are completely 
uncorrelated. 
The data for $r$ is not at all surprising as we already observed that even for consecutive years, correlation is not large.

\begin{figure}
\begin{center}
\includegraphics[width=11.0cm]{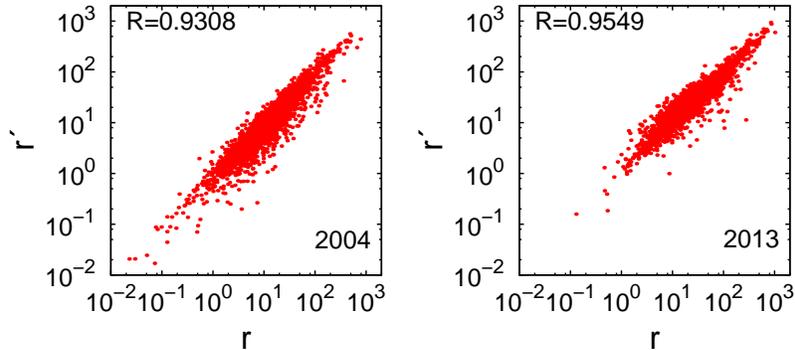}
     \caption{
      Correlation of $r$ and $r^{\prime}$ for 2004 ($R=0.9308$) and 2013 
($R=0.9549$).
}
\label{fig:cr_cr_corr}
\end{center}
 \end{figure}

 The correlation between $r$ and  $r^{\prime}$ 
is found to be quite high. In Fig.~\ref{fig:cr_cr_corr}, we show the 
correlation between $r$ and $r^{\prime}$ for two years, $R=0.9308$ for 2004 
and $R=0.9549$ for 2013.



\begin{table}[h]
\begin{center}
\caption{Table for the value of the correlation coefficient $R$ for $r$ for all the pairs of consecutive  years.}
\begin{tabular}{| c| c |}
\hline
Pairs of year    &       $ R $         \\
\hline
 $2004-2005$ & $0.7923 $  \\
 $2005-2006$ & $0.8680$  \\
 $2006-2007$ & $0.8177$  \\
 $2007-2008$ & $0.7543 $ \\
 $2008-2009$ & $0.9426$   \\
 $2009-2010$ & $0.9039$   \\
 $2010-2011$ & $0.9040$   \\
 $2011-2012$ & $0.9297$  \\
 $2012-2013$ & $0.9327$   \\
 \hline
\end{tabular}
\label{tab:coefficient}
\end{center}
\end{table}

\subsection{Distribution of  annual citations, IF and annual citation rate: nature of their tails}
\label{sec:tails}

\begin{figure}
\begin{center}
\includegraphics[width=11.0cm]{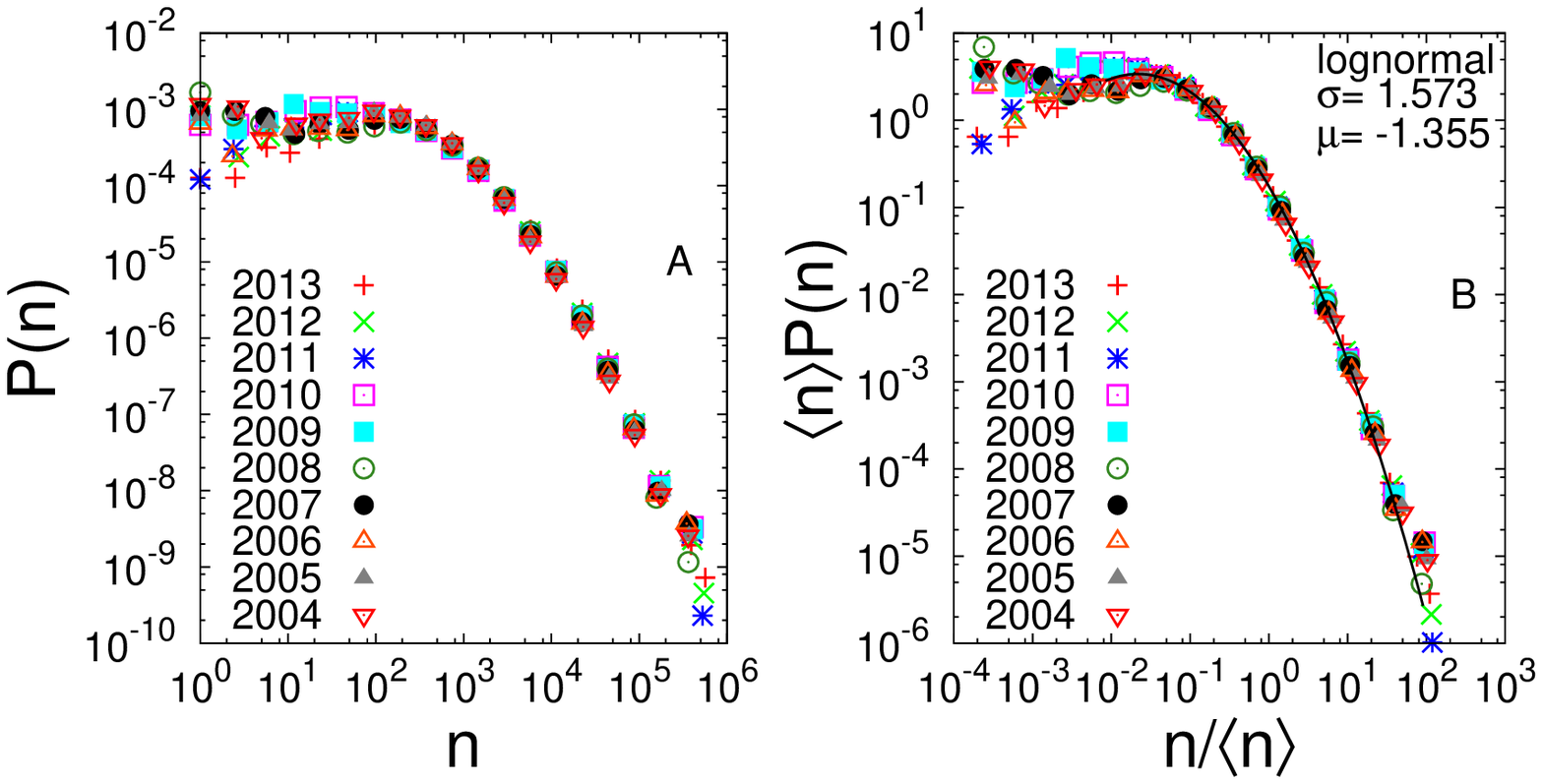}
\includegraphics[width=11.0cm]{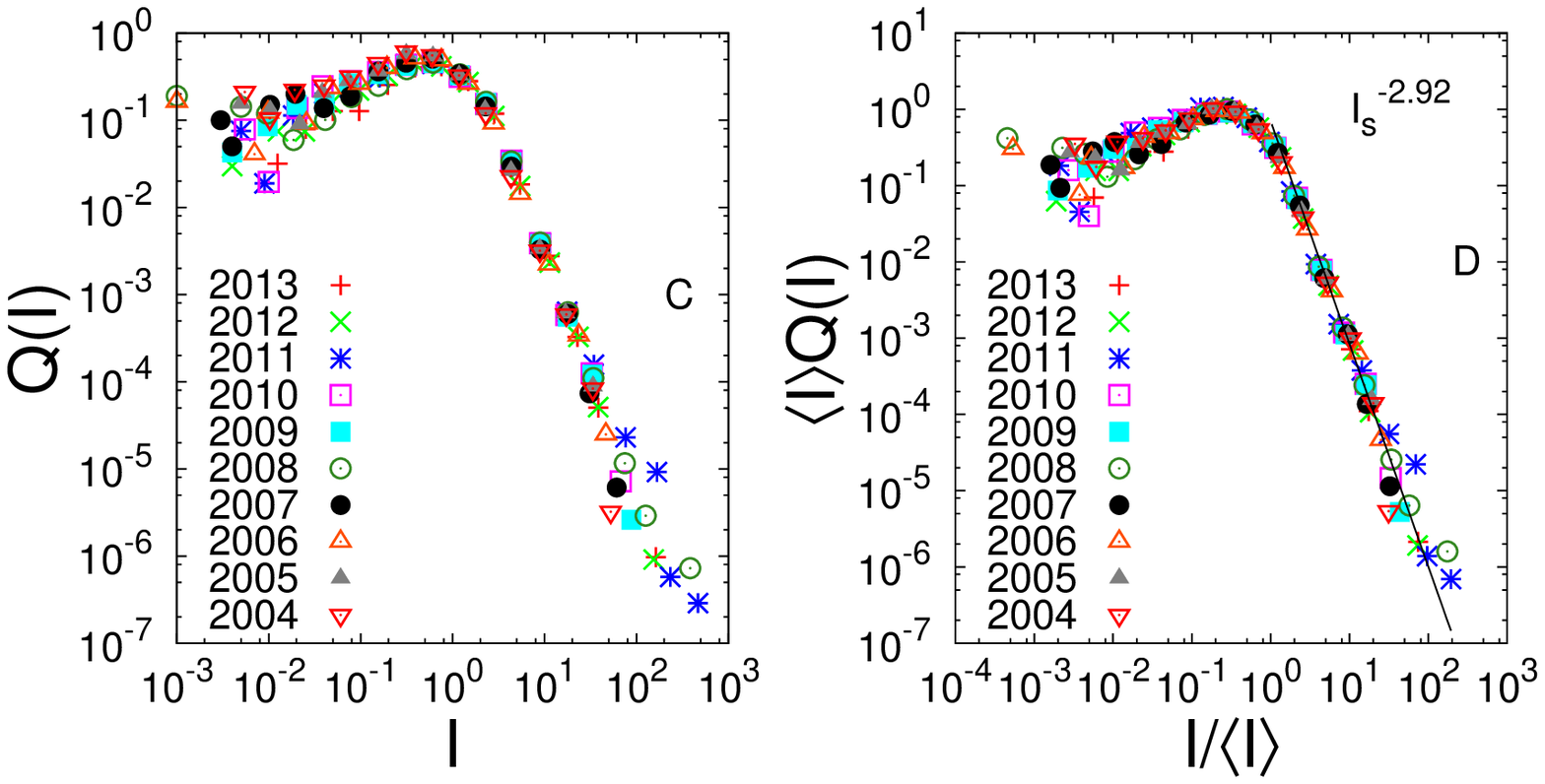}
     \caption{
     (A) Probability distribution of annual citations $P(n)$ and 
   (B) scaling collapse of the same, which 
   fits fairly well to a 
   lognormal form.
  (C) Probability distribution of impact factor $Q(I)$ and 
 (D) scaling collapse of the same, the tail
 fits fairly well to a power law form.
The straight line has slope $\gamma_I = 2.92$.
}
\label{fig:dist}
\end{center}
 \end{figure}

First we  investigate  the nature of the tail of the distribution of
annual citations $P(n)$ (Fig.~\ref{fig:dist}(A),(B)) and  impact factors $Q(I)$ 
(Fig.~\ref{fig:dist}(C),(D)).
The plots showed excellent scaling collapse over years when 
in general for any  probability distribution $X(x)$, $X(x) \langle x\rangle$ is
plotted against $x / \langle x \rangle$.
The distribution of annual citations and  impact factors show non-monotonic 
behavior, with a peak occurring approximately at half the average values.
The tail of the annual citations distributions 
(Fig.~\ref{fig:dist}(B))
fit well to
a lognormal form 
($X(x)=\frac{1}{x\sigma\sqrt{2\pi}} e ^{-\frac {(\log x-\mu)^2}{2\sigma^2}}$)
with $\mu=-1.355$ and $\sigma = 1.573$.
However, the tail of the impact factor distribution fits to a power law,
with a decay exponent about $\gamma_I=2.92$.

\begin{figure}
\begin{center}
\includegraphics[width=11.0cm]{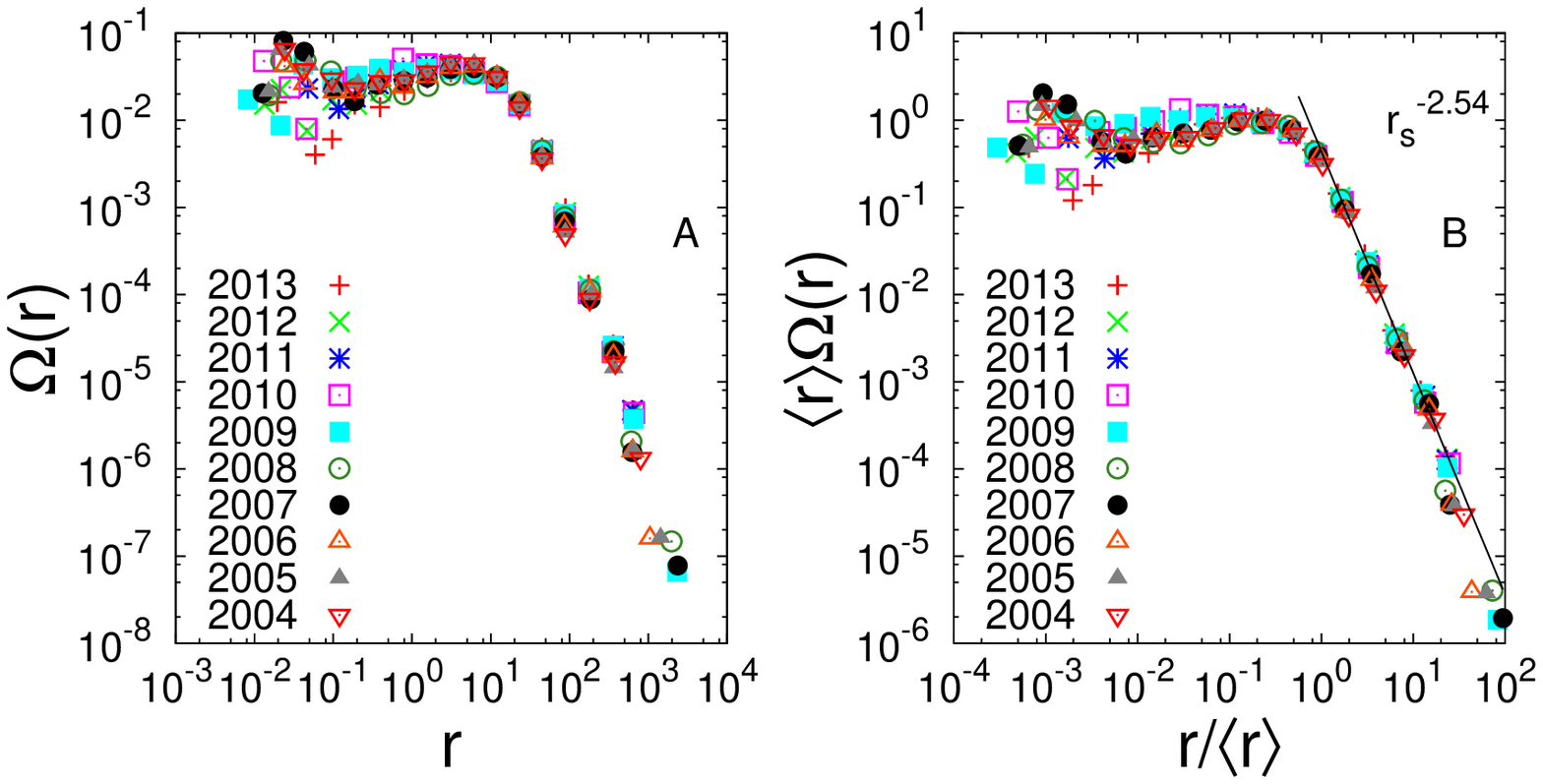}
\includegraphics[width=11.0cm]{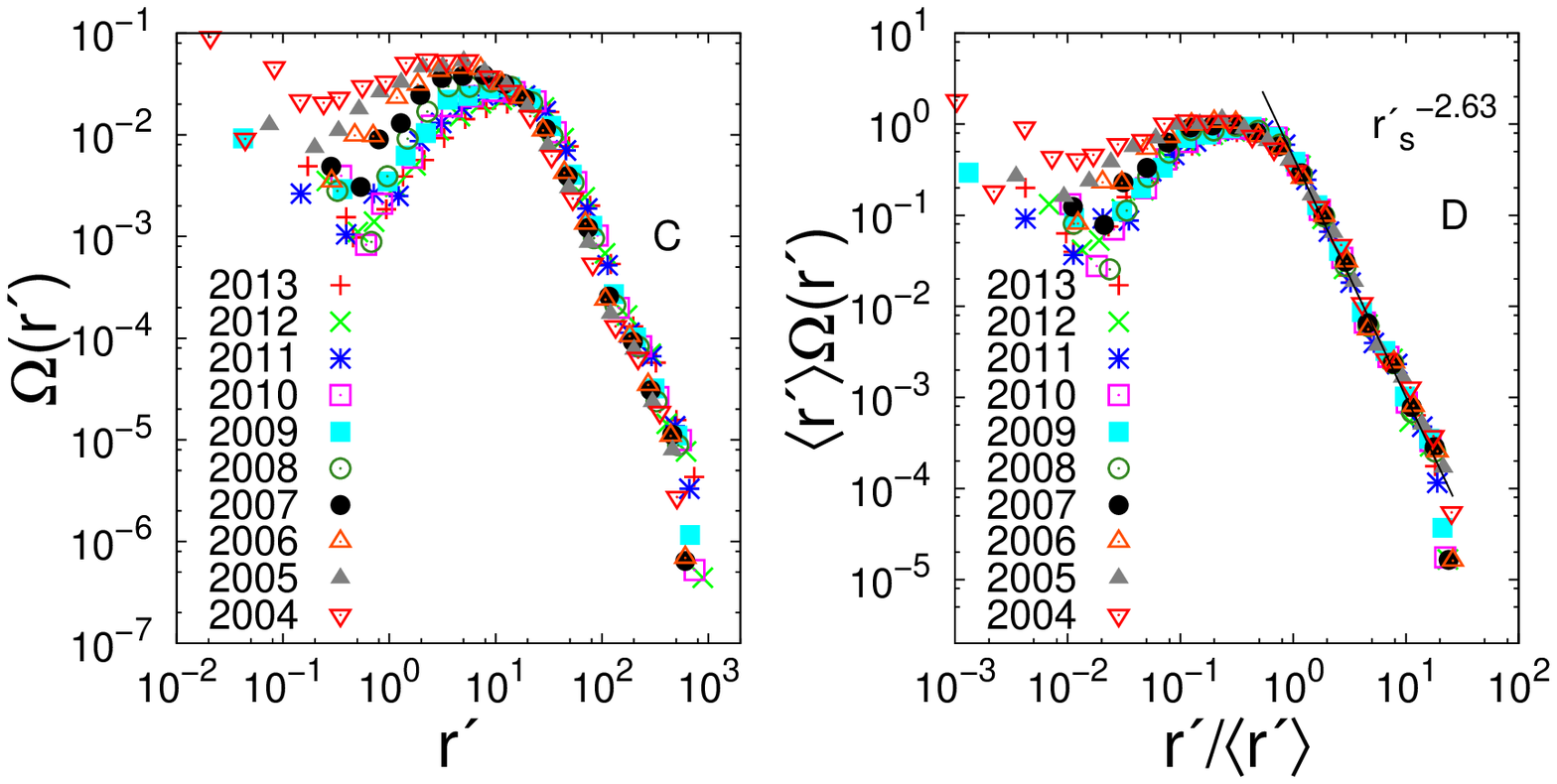}
     \caption{
     (A) Probability distribution of annual citation rates $\Omega(r)$ and 
   (B) scaling collapse of the same, the tail fits fairly well to a 
   power law form with a decay exponent $\gamma_r=2.54$.
 (C) Probability distribution of annual citation rates 
$\Omega(r^{\prime})$ and (D) scaling collapse of the same, the tail
   fits to a power law form with a decay exponent $\gamma_r=2.63$. 
   $r^{\prime}$ being calculated using the average number of publications for 
the last 10 years.
}
\label{fig:rdist}
\end{center}
 \end{figure}

The probability
distributions $\Omega(r)$  of the newly proposed quantity, the annual citation 
rate $r$ also shows a power law tail with a decay exponent $\gamma_r \approx 
2.54$ (Fig~\ref{fig:rdist}(B)); although it is almost a flat distribution for 
$r/\langle r \rangle < 1$. 
The probability distribution $\Omega(r^{\prime})$ 
shows similar features as $\Omega(r)$ and its tail resembles roughly a power 
law with decay exponent  $\gamma_r{\prime} \approx 2.63$ (Fig~\ref{fig:rdist}(D)).
In fact, we checked the correlation between $r$ and $r^{\prime}$ for each year,
and found them to be very strongly correlated ($R > 0.9$).

If $I$ and $r$ are related by a power law, one can in principle derive the
exponent of the distribution of $r$ from that of $I$.
Assuming in general the scaled distribution of $I$ has  a power law tail with 
exponent $\gamma_I $ and
$I \propto r^{\xi_r}$, the tail of the distribution for the scaled $r$ should 
follow the behaviour
$(r/\langle r \rangle)^{-\gamma_I {\xi_r} + {\xi_r} -1}$.
However, we have noted earlier that $\xi_r$ is not unique.
Putting the value of $\gamma_I = 2.92$ and the observed value $\gamma_r = 2.54$,
we obtain $\xi_r \approx 0.80$. It is interesting to note that this
 value does not correspond to either of the two values of $\xi_r$
(see Table.~\ref{tab:expoir}) estimated from the $I-r$ curves but rather
 is very close to the average of the two values in general.

%
\section{Summary  and discussions}

We analyzed the citation data to Science journals for the period 2004-13 
considering the entire data set available in the Journal Citation 
Reports~\cite{JCR}.
The analysis is based on four  different measures or indices. While the 
impact factor $I$ and annual citation ($n$)  are readily available, we also introduced a third index, the  citation rate $r$,
which can be  easily estimated from the  database. Closely related to $r$ is another measure $r\prime$ which is calculated combining
 different years.

We have primarily studied the correlation between different measures in the same year, correlation between the same measures
in two different years and probability distribution functions for different 
years.
In this paper, we reported the explicit functional forms by  
which any pair of the three -- $I$, $n$ and $r$ are related to each other. Also correlation between $r$ and $r\prime$ has been studied.
The most surprising result is perhaps the fairly strong dependence of $I$ on $n$.
This dependence is not accidental  as the corresponding exponent does not show appreciable change over time. It is to be remembered that 
$I$ depends on the citations to recent publications only while $n$ considers citations to all 
published papers in a journal. One might expect that old journals will have larger value of $n$ as a result.
Impact factor $I$ on the other hand will not depend on the age of the journal, 
in principle. Thus it is quite surprising to see that $I$
 and $n$ show a strong functional dependence. The measure $r$ is also expected to be dependent on the age.
However, in contrast to $n$ it is a scaled data as $I$, albeit in a different way, 
 and it is not surprising that $I$ depends on $r$ in a stronger manner.  However $r$ being quite
 arbitrarily defined, the correlation between $I$ and $r$ is not apparent. In fact, the numerator and denominator of $r$
  are completely independent variables unless the number of publications is same for all years which is not the case.
  For  $I$ on the other hand, the quantities appearing in the numerator and denominator are directly correlated. 
  This work therefore opens up the question whether the impact factor is the most reliable measure or not. If so, then the present work indicates that 
  $n$ and $r$ might as well qualify as such measures, which is less than obvious.



The probability distributions of the three measures are found to show 
conventional behavior, i.e., either log-normal type or they occur with power law 
tails.
Like many other real systems, the exponents for power law variation lie between $2$ and $3$ which is
sensible so as to have a finite average value.
The scaled distribution of different years show universal behavior in time, as 
has been observed in many other systems in which the 
dynamics of popularity has been studied~\cite{chatterjee2013universality}.
The preference of lognormals and power laws are decided from
eye estimates of best fits. However, the reason why annual citations may follow
a lognormal distribution can be justified by the fact that annual citations are 
very highly correlated, and one can imagine an underlying multiplicative 
process.

\acknowledgements The authors thank A. Ghosh for assisting in data acquisition.
AK acknowledges financial support from UGC sanction No. F.7-48/2007 (BSR).
PS acknowledges financial support from CSIR project.
{}

\end{document}